\documentclass[a4paper,aps,prd,10pt,preprintnumbers,showpacs,twocolumn,superscriptaddress,nofootinbib,amsmath,amssymb]{revtex4-1}
\usepackage[dvips]{graphics}
\usepackage{cmap}
\usepackage[utf8]{inputenc}
\usepackage[T1]{fontenc}

\def\imo{i}

\def\K{{\cal K}}

\begin{document}
\title{Quantum gravitational corrections to the Schwarzschild spacetime and quasinormal frequencies}
\author{Alexey Dubinsky}\email{dubinsky@ukr.net}
\affiliation{University of Sevilla, 41009 Seville, Spain}
\begin{abstract}
Quantum gravitational corrections to the entropy of the Schwarzschild black hole, derived using the Wald entropy formula within an effective field theory framework, were presented in [X. Calmet, F. Kuipers Phys.Rev.D 104 (2021) 6, 066012]. These corrections result in a Schwarzschild spacetime that is deformed by the quantum correction. However, it is observed that the proposed quantum-corrected metric describes not a black hole, but a wormhole. Nevertheless, further expansion of the metric function in terms of the quantum correction parameter yields a well-defined black hole metric whose geometry closely resembles that of a wormhole. We also explore methods for distinguishing between these quantum-corrected spacetimes based on the quasinormal frequencies they emit. We show that while the fundamental mode deviates from the Schwarzschild limit only mildly, the first few overtones deviate at a strongly increasing rate, creating a characteristic ``sound'' of the event horizon.
\end{abstract}
\maketitle
\section{Introduction}

Observations of quasinormal modes of black holes offer valuable insights into their fundamental characteristics. By detecting these characteristic vibrational frequencies in gravitational wave signals emitted during events such as black hole mergers, scientists can directly investigate the spacetime geometry surrounding these cosmic entities, providing information on their mass, spin, and potential departures from classical predictions \cite{LIGOScientific:2016aoc,LIGOScientific:2017vwq,LIGOScientific:2020zkf,Babak:2017tow}. Such observations serve as a critical tool for validating theoretical models of black hole physics and advancing our understanding of the universe's most intriguing phenomena.

Various theories of gravity aim to develop a quantum theory of gravity or to introduce quantum corrections to the classical solutions of Einstein's gravity. In pursuit of this goal, numerous efforts have been made to construct models of Schwarzschild-like black holes with quantum corrections. Perturbations, scattering properties, and the quasinormal spectrum of black hole geometries with such quantum corrections have been extensively studied across various approaches, as documented in prior research \cite{Konoplya:2023ahd,Moreira:2023cxy,Liu:2020ola,Xing:2022emg,Fu:2023drp,Yang:2022btw,Karmakar:2022idu,Cruz:2020emz,Saleh:2014uca,Liu:2012ee}. Recently, Calmet and Kuipers \cite{Calmet:2021lny} developed a model of a quantum-corrected spacetime, calculating quantum gravitational corrections to black hole entropy using the Wald entropy formula within an effective field theory framework. These corrections, extended to the second order in curvature and a subset of the third order, were found to induce adjustments to the horizon radius and temperature, as detailed in their work.

However, to our knowledge, neither the quasinormal spectrum of the Calmet-Kuipers spacetime \cite{Calmet:2021lny} nor its properties have been thoroughly investigated. Our objective here is to analyze the primary characteristics of this quantum-corrected spacetime and determine methods for distinguishing it from the Schwarzschild metric based on their respective quasinormal spectra.

We will demonstrate that the Calmet-Kuipers spacetime actually describes a wormhole rather than a black hole. However, by retaining only the linear correction term in one of the metric functions, it is possible to transform this spacetime into a black hole metric that closely resembles the original wormhole geometry. Additionally, we will examine the quasinormal modes of scalar, neutrino, and electromagnetic perturbations in this quantum-corrected black hole spacetime. Our findings will reveal that while the real oscillation frequency, determined by the real part of the complex quasinormal mode, is minimally affected by the quantum correction, the damping rate exhibits a soft yet noticeable decrease due to the quantum correction.

The paper is structured as follows: In Section \ref{sec:wavelike}, we provide an overview of the quantum-corrected metric and demonstrate its characterization of a wormhole spacetime. Subsequently, we introduce the corresponding black hole metric and explore the wave-like equations and effective potentials in Section \ref{sec:methods}. Section \ref{sec:QNM} delves into the computation of quasinormal modes, detailing the three methods utilized: Frobenius or Leaver method, time-domain integration and the WKB approach. Finally, in Section \ref{sec:conclusions}, we offer a summary of the findings obtained.

\section{Higher curvature corrections}\label{sec:wavelike}

In \cite{Calmet:2021lny} the analysis starts from the effective action to quantum gravity~\cite{Weinberg:1980gg,Barvinsky:1983vpp,Barvinsky:1985an,Barvinsky:1987uw,Donoghue:1994dn}. At second order in curvature, it was found that
\begin{widetext}
\begin{align}\label{EFTaction}
S_{\text{EFT}} = \int \sqrt{|g|}d^4x  \left( \frac{R}{16\pi G_N} + c_1(\mu) R^2 + c_2(\mu) R_{\mu\nu} R^{\mu\nu} + c_3(\mu) R_{\mu\nu\rho\sigma} R^{\mu\nu\rho\sigma} + \mathcal{L}_m \right) \ ,
\end{align}
for the local part of the action and the nonlocal part is given by
\begin{align}\label{nonlocalaction}
	\Gamma_{\text{NL}}^{\scriptstyle{(2)}}  = - \int  \sqrt{|g|}d^4x \left[ \alpha R \ln\left(\frac{\Box}{\mu^2}\right)R + \beta R_{\mu\nu} \ln\left(\frac{\Box}{\mu^2}\right) R^{\mu\nu} + \tilde{\gamma} R_{\mu\nu\alpha\beta} \ln\left(\frac{\Box}{\mu^2}\right)R^{\mu\nu\alpha\beta} \right],
	\end{align}
	where $\Box := g^{\mu\nu} \nabla_\mu \nabla_\nu$.
\end{widetext}
There are no correction to the Wald formula \cite{Wald:1993nt} at the second order, so that the third order in curvature correction was taken into account in  \cite{Calmet:2021lny},
\begin{eqnarray}
{\cal L}^{(3)}=c_6 G_N  R^{\mu\nu}_{\ \ \alpha\sigma} R^{\alpha\sigma}_{\ \ \delta\gamma} R^{\delta\gamma}_{\ \ \mu\nu}  \ ,
\end{eqnarray}
where $c_6$ is dimensionless.

The above thrid order correction leads to the Calmet-Kuipers metric  \cite{Calmet:2021lny},
\begin{equation}\label{metric}
  ds^2=-f(r)dt^2+\frac{B^2(r)}{f(r)}dr^2+r^2(d\theta^2+\sin^2\theta d\phi^2),
\end{equation}
where
$$
\begin{array}{rcl}
f(r)&=&\displaystyle 1+ \frac{5 \gamma  M^3}{r^7}-\frac{2 M}{r},\\
B^{2}(r)&=&\displaystyle \frac{1+\frac{5 \gamma  M^3}{r^7}-\frac{2 M}{r}}{1+ \frac{\gamma  M^2 \left(27-\frac{49 M}{r}\right)}{r^6}-\frac{2 M}{r}},\\
\end{array}
$$
where $\gamma = 128 \pi c_{6} G_{N}^5$.

\begin{figure*}
\resizebox{\linewidth}{!}{\includegraphics{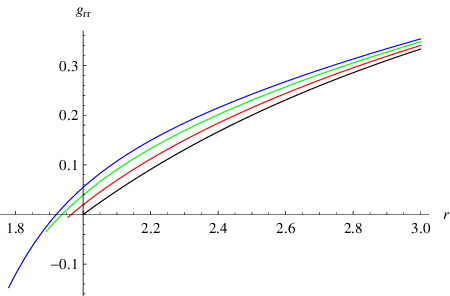}\includegraphics{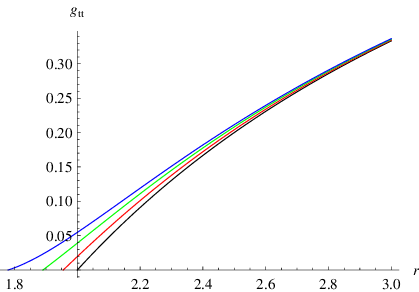}\includegraphics{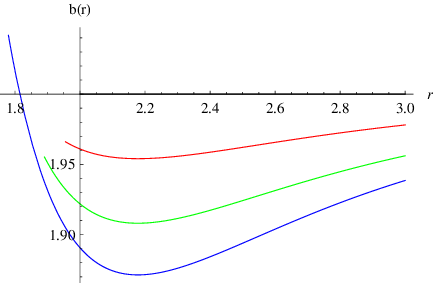}}
\caption{The metric function $g_{rr}(r)$ (left) and $g_{tt}$ (middle), together with the shape function $b(r)$, starting from zero of $g_{tt}$, $M=1$: $\gamma=0$ (black, bottom), $\gamma=0.5$ (red), $\gamma=1$ (green), $\gamma=1.4$ (blue, top).}\label{fig:metric}
\end{figure*}

It is stated in \cite{Calmet:2021lny} that the above metric leads to the correction of the event horizon
\begin{eqnarray}\label{horizon}
r_H= 2 G_N M \left (1 - c_6 \frac{5 \pi}{G_N^2 M^4 }\right).
\end{eqnarray}
However, the above corrected metric does not correspond to a black hole, because the zero of the metric function $f(r)$ corresponds to the negative value of $B^2/f$ as can be seen in fig. \ref{fig:metric}.

%%%%%%%%%%%%%%%%%%%%%%%%%
Static, spherically symmetric, Lorentzian traversable wormholes of arbitrary geometrical configuration can be effectively described using the Morris-Thorne framework \cite{Morris:1988cz}. The corresponding spacetime metric is expressed as follows:
\begin{equation}\label{MT}
ds^2 = - e^{2 \Phi (r)} dt^2 + \frac{dr^2}{1 - \frac{b(r)}{r}} + r^2 (d\theta^2 + \sin^2\theta \, d\phi^2),
\end{equation}
where the function \(\Phi(r)\) represents the lapse function, which dictates both the gravitational redshift and the tidal forces present in the wormhole's spacetime. Specifically, when \(\Phi = 0\), the wormhole does not exert any tidal forces on objects traversing it. The spatial structure or shape of the wormhole is entirely determined by another function, \(b(r)\), referred to as the shape function.

When one embeds the wormhole metric into a Euclidean space-time using cylindrical coordinates, the geometry of the embedded surface within the equatorial plane (\(\theta = \pi/2\)) is governed by the differential equation:
\begin{equation}
\frac{dz}{dr} = \pm \left(\frac{r}{b(r)} - 1\right)^{-\frac{1}{2}},
\end{equation}
The throat of the wormhole, which is the narrowest part, corresponds to the minimum value of the radial coordinate \(r\), denoted as \(r_{\text{min}} = b_0\). Consequently, the coordinate \(r\) ranges from \(r_{\text{min}}\) out to spatial infinity \(r = \infty\). However, when considering the proper radial distance \(dl\), which is given by:
\begin{equation}
\frac{dl}{dr} = \pm \left(1 - \frac{b(r)}{r}\right)^{-\frac{1}{2}},
\end{equation}
the wormhole structure is such that the radial distance extends to two infinities, \(l = \pm \infty\), at \(r = \infty\).

For the wormhole spacetime to be regular, the lapse function \(\Phi(r)\) must remain finite at all points. Additionally, asymptotic flatness requires that \(\Phi(r) \rightarrow 0\) as \(r \rightarrow \infty\) (or equivalently as \(l \rightarrow \pm \infty\)). The shape function \(b(r)\) must satisfy the conditions \(1 - \frac{b(r)}{r} > 0\) and \(\frac{b(r)}{r} \rightarrow 0\) as \(r \rightarrow \infty\) (or \(l \rightarrow \pm \infty\)). At the throat, where \(r = b(r)\), the expression \(1 - \frac{b(r)}{r}\) approaches zero, ensuring that the metric remains nonsingular at this critical juncture, allowing a traveler to pass through the wormhole in a finite amount of time. It is worth mentioning that the quasinormal frequencies for the above wormholes could be found in a manner similar to \cite{Konoplya:2010kv} and the boundary conditions are the same as for the black hole in terms of the tortoise coordinate \cite{Konoplya:2005et}.

%%%%%%%%%%%%%%%%%%%%%%%%%%%%%%%

If one introduces the shape function $b(r)$ as follows
\begin{equation}
b(r) = \left(1- \frac{f(r)}{B^2(r)}\right) r,
\end{equation}
then, from fig.  \ref{fig:metric} we see that the this function has a minimum which is always larger than the zero of the function $f(r)$.
This minimum denotes the throat of a wormhole. However, if we further expand the metric function $B(r)$ keeping only the linear order in $\gamma$ and neglecting higher orders, we obtain a black hole metric whose geometry is quite close to that of the black hole once $\gamma$ is sufficiently small.

This way, we obtain
$$
\begin{array}{rcl}
f(r)&=&\displaystyle 1+ \frac{5 \gamma  M^3}{r^7}-\frac{2 M}{r},\\
B(r)&=&\displaystyle 1-\frac{27 \gamma  M^2}{2 r^6},\\
\end{array}
$$
where $\gamma $ is the quantum parameter, and $M$ is the ADM mass. We shall further measure all dimensional quantities in units of the mass, i.~e., we choose $M=1$.
This black hole metric is compatible with the correction to the event horizon given by eq. (\ref{horizon}).

The general relativistic equations for the scalar ($\Phi$), electromagnetic ($A_\mu$), and Dirac ($\Upsilon$) fields can be written in the following form:
\begin{subequations}\label{coveqs}
\begin{eqnarray}\label{KGg}
\frac{1}{\sqrt{-g}}\partial_\mu \left(\sqrt{-g}g^{\mu \nu}\partial_\nu\Phi\right)&=&0,
\\\label{EmagEq}
\frac{1}{\sqrt{-g}}\partial_{\mu} \left(F_{\rho\sigma}g^{\rho \nu}g^{\sigma \mu}\sqrt{-g}\right)&=&0\,,
\\\label{covdirac}
\gamma^{\alpha} \left( \frac{\partial}{\partial x^{\alpha}} - \Gamma_{\alpha} \right) \Upsilon&=&0,
\end{eqnarray}
\end{subequations}
where $F_{\mu\nu}=\partial_\mu A_\nu-\partial_\nu A_\mu$ is the electromagnetic tensor, $\gamma^{\alpha}$ are noncommutative gamma matrices and $\Gamma_{\alpha}$ are spin connections associated with the tetrads.
The above equations (\ref{coveqs}) can be reduced the following wavelike form \cite{Kokkotas:1999bd,Berti:2009kk,Konoplya:2011qq}:
\begin{equation}\label{wave-equation}
\dfrac{d^2 \Psi}{dr_*^2}+(\omega^2-V(r))\Psi=0,
\end{equation}
where the ``tortoise coordinate'' $r_*$ is
\begin{equation}\label{tortoise}
dr_*\equiv\frac{B(r)}{f(r)}dr.
\end{equation}

The effective potentials for the scalar ($s=0$) and electromagnetic ($s=1$) fields are expressed as follows
\begin{equation}\label{potentialScalar}
V(r)=f(r)\frac{\ell(\ell+1)}{r^2}+\frac{1-s}{r}\cdot\frac{d^2 r}{dr_*^2},
\end{equation}
where $\ell=s, s+1, s+2, \ldots$ are the multipole numbers.
The Dirac field ($s=1/2$) has two isospectral potentials,
\begin{equation}
V_{\pm}(r) = W^2\pm\frac{dW}{dr_*}, \quad W\equiv \left(\ell+\frac{1}{2}\right)\frac{\sqrt{f(r)}}{r}.
\end{equation}
The isospectral wave functions can be transformed one into another by the Darboux transformation,
\begin{equation}\label{psi}
\Psi_{+}\propto \left(W+\dfrac{d}{dr_*}\right) \Psi_{-},
\end{equation}
so that only one of the effective potentials, and here we choose $V_{+}(r)$, is sufficient for quasinormal modes analysis.

\begin{figure}
\resizebox{\linewidth}{!}{\includegraphics{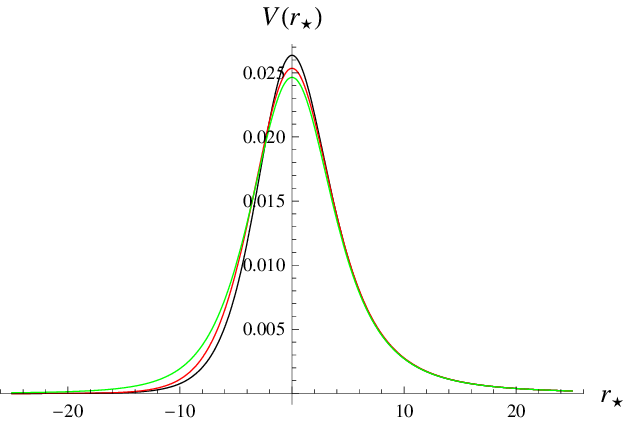}}
\caption{Potential as a function of the tortoise coordinate of the $\ell=0$ scalar field for the Calmet-Kuipers black hole ($M=1$): $\gamma=0$ (black), $\gamma=1$ (red), $\gamma=1$ (green).}\label{fig:potentials}
\end{figure}

\begin{figure}
\resizebox{\linewidth}{!}{\includegraphics{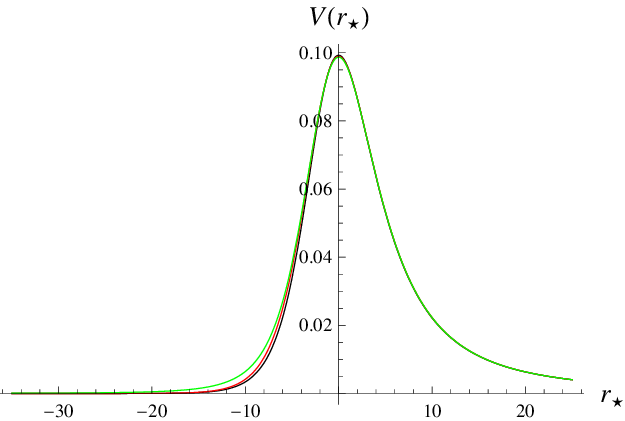}}
\caption{Potential as a function of the tortoise coordinate of the $\ell=1$ scalar field for the Calmet-Kuipers black hole ($M=1$): $\gamma=0$ (black), $\gamma=1$ (red), $\gamma=1.45$ (green).}\label{fig:potentials2}
\end{figure}

\begin{figure}
\resizebox{\linewidth}{!}{\includegraphics{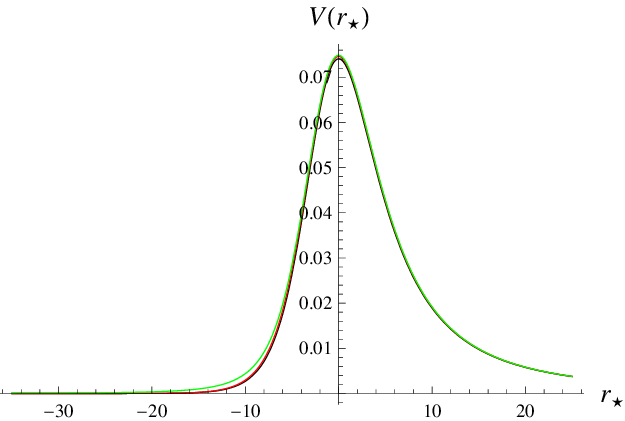}}
\caption{Potential as a function of the tortoise coordinate of the $\ell=1$ scalar field for the Calmet-Kuipers black hole ($M=1$): $\gamma=0$ (black), $\gamma=1$ (red), $\gamma=1.45$ (green).}\label{fig:potentialsEM}
\end{figure}

\begin{figure}
\resizebox{\linewidth}{!}{\includegraphics{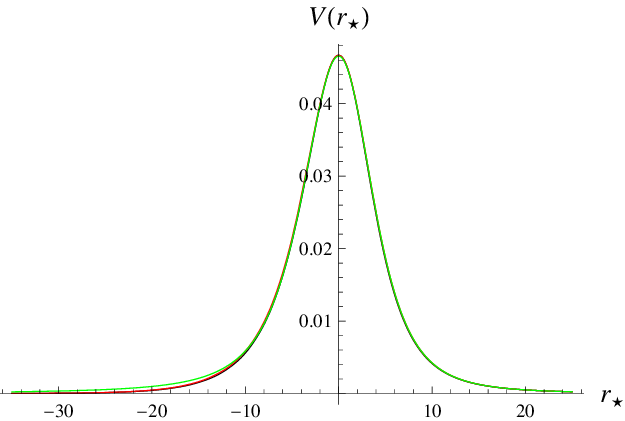}}
\caption{Potential as a function of the tortoise coordinate of the $\ell=1/2$ scalar field for the Calmet-Kuipers black hole ($M=1$, $\Lambda=1/10$): $\gamma=0$ (black), $\gamma=1$ (red), $\gamma=1.45$ (green).}\label{fig:potentials3}
\end{figure}

Effective potentials for the scalar, electromagnetic and Dirac fields are given in figs. \ref{fig:potentials}-\ref{fig:potentials3}.
The potentials are only slightly corrected by the quantum parameter. Therefore, one should  not have large correction to the fundamental quasinormal frequency, as it is expected from the quantum correction.

\section{The methods}\label{sec:methods}

Quasinormal modes of asymptotically flat black holes satisfy the following boundary conditions
\begin{equation}\label{boundaryconditions}
\Psi(r_*\to\pm\infty)\propto e^{\pm\imo \omega r_*},
\end{equation}
which represent purely ingoing waves at the horizon ($r_*\to-\infty$) and purely outgoing waves at spatial infinity ($r_*\to\infty$).

\subsection{Time-domain integration}

As the time-domain integration, we applied the Gundlach-Price-Pullin discretization scheme \cite{Gundlach:1993tp}, represented as follows:
\begin{eqnarray}
\Psi\left(N\right)&=&\Psi\left(W\right)+\Psi\left(E\right)-\Psi\left(S\right)\nonumber\\
&&- \Delta^2V\left(S\right)\frac{\Psi\left(W\right)+\Psi\left(E\right)}{8}+{\cal O}\left(\Delta^4\right),\label{Discretization}
\end{eqnarray}
Here, the integration scheme encompasses points denoted as $N\equiv\left(u+\Delta,v+\Delta\right)$, $W\equiv\left(u+\Delta,v\right)$, $E\equiv\left(u,v+\Delta\right)$, and $S\equiv\left(u,v\right)$. This discretization method has been employed in various studies \cite{Konoplya:2014lha,Konoplya:2020jgt,Konoplya:2005et,Varghese:2011ku,Momennia:2022tug,Qian:2022kaq}, demonstrating its reliability.

To extract frequency values from the time-domain profile, we employ the Prony method, which entails fitting the profile data with a sum of damped exponents (see for instance \cite{Konoplya:2011qq} for details). Quasinormal modes are typically derived from time-domain profiles when the ring-down stage encompasses a sufficient number of oscillations. The duration of the ringdown period increases with the multipole number $\ell$.

%\begin{equation}\label{damping-exponents}
%\Psi(t)\simeq\sum_{i=1}^pC_ie^{-i\omega_i t}.
%\end{equation}
%We assume that the quasinormal ringing phase initiates at $t_0=0$ and terminates at $t=Nh$, where $N\geq2p-1$. Consequently, relation (\ref{damping-exponents}) holds for each %profile point:
%\begin{equation}
%x_n\equiv\Psi(nh)=\sum_{j=1}^pC_je^{-i\omega_j nh}=\sum_{j=1}^pC_jz_j^n.
%\end{equation}
%Subsequently, we ascertain $z_i$ based on the known $x_n$ and compute the quasinormal frequencies $\omega_i$. Quasinormal modes are typically derived from time-domain profiles %when the ring-down stage encompasses a sufficient number of oscillations. Moreover, the duration of the ringdown period increases with the multipole number $\ell$.

\begin{table*}
\begin{tabular}{c c c c}
\hline
$\gamma $ & Frobenius & time-domain & WKB6 Padé  \\
\hline
$0$ & $0.110455 - 0.104896 i$ &$0.110445-0.105084 i$ & $0.110792-0.104683 i$ \\
$0.25$ & $0.111313 - 0.104345 i$ &$0.111296-0.104531 i$ & $0.109935-0.102888 i$ \\
$0.5$ & $0.112266 - 0.103602 i$ &$0.112238-0.103785 i$ & $0.109468-0.099549 i$ \\
$0.75$ & $0.113327 - 0.102546 i$ &$0.113281-0.102727 i$ & $0.106818-0.105542 i$ \\
$1.$ & $0.114482 - 0.10091 i$ & $0.114405-0.101091 i$ & $0.107774-0.105685 i$ \\
$1.25$ & $0.115432 - 0.0978777 i$ &$0.115312-0.098050 i$ & $0.119765-0.087833 i$ \\
$1.45$ & $0.111954 - 0.0945804 i$ &$0.111866-0.094720 i$ & $0.111334-0.096194 i$\\
\hline
\end{tabular}
\caption{Comparison of the quasinormal frequencies obtained by the Frobenius method, time-domain integration and the 6th order WKB approach with Padé approximants for $s=\ell=0$ ($M=1$).}\label{check1}
\end{table*}

\begin{table*}
\begin{tabular}{c c c c}
\hline
$\gamma $ & Frobenius & time-domain & WKB6 Padé  \\
\hline
$0$ & $0.292936 - 0.09766 i$ & $0.292945-0.097663 i$ & $0.292930-0.097663 i$ \\
$0.25$ & $0.29325 - 0.0972036 i$ & $0.293260-0.097206 i$ & $0.293220-0.097256 i$ \\
$0.5$ & $0.293514 - 0.0966556 i$  & $0.293524-0.096658 i$ & $0.293595-0.096750 i$\\
$0.75$ & $0.293676 - 0.0959935 i$  & $0.293686-0.095996 i$ & $0.293808-0.095906 i$\\
$1.$ &  $0.293619 - 0.095208 i$   & $0.293630-0.095210 i$ & $0.293507-0.095045 i$ \\
$1.25$ & $0.293056 - 0.0944508 i$ & $0.293067-0.094451 i$ & $0.292756-0.094573 i$ \\
$1.45$ & $0.292469 - 0.0946128 i$ & $0.292479-0.094612 i$ & $0.292446-0.094675 i$\\
\hline
\end{tabular}
\caption{Comparison of the quasinormal frequencies obtained by the Frobenius method, time-domain integration and the 6th order WKB approach with Padé approximants for $s=0$, $\ell=1$ ($M=1$).}\label{check2}
\end{table*}

\begin{table*}
\begin{tabular}{c c c c c}
\hline
$\gamma $ & time-domain & WKB6 Padé & rel. diff. $Re (\omega)$ & rel. diff. $Im (\omega)$ \\
\hline
$0$ & $0.183031-0.096907 i$ & $0.182643-0.096566 i$ & $0.212\%$ & $0.352\%$\\
$0.25$ & $0.183407-0.096792 i$ & $0.183384-0.097017 i$ & $0.0127\%$ & $0.233\%$\\
$0.5$ & $0.183759-0.096612 i$ & $0.184060-0.096731 i$ & $0.164\%$ & $0.123\%$\\
$0.75$ & $0.184035-0.096346 i$ & $0.185120-0.095946 i$ & $0.590\%$ & $0.415\%$\\
$1.$ & $0.184089-0.095982 i$ & $0.181304-0.093708 i$ & $1.51\%$ & $2.37\%$\\
$1.25$ & $0.183407-0.095771 i$ & $0.181761-0.096106 i$ & $0.897\%$ & $0.350\%$\\
$1.45$ & $0.183622-0.097793 i$ & $0.183519-0.096814 i$ & $0.0561\%$ & $1.00\%$\\
\hline
\end{tabular}
\caption{Comparison of the quasinormal frequencies obtained by the time-domain integration and the 6th order WKB approach with Padé approximants for $s=1/2$, $\ell=1/2$ ($M=1$).}\label{check3}
\end{table*}

\begin{table*}
\begin{tabular}{c c c c c}
\hline
$\gamma $ & time-domain & WKB6 Padé & rel. diff. $Re (\omega)$ & rel. diff. $Im (\omega)$ \\
\hline
$0$ & $0.380042-0.096388 i$ & $0.380041-0.096408 i$ & $0.00015\%$ & $0.0207\%$\\
$0.25$ & $0.380497-0.096181 i$ & $0.380495-0.096204 i$ & $0.00057\%$ & $0.0239\%$\\
$0.5$ & $0.380937-0.095940 i$ & $0.380932-0.095971 i$ & $0.00137\%$ & $0.0326\%$\\
$0.75$ & $0.381343-0.095668 i$ & $0.381334-0.095706 i$ & $0.00235\%$ & $0.0390\%$\\
$1.$ & $0.381689-0.095395 i$ & $0.381677-0.095450 i$ & $0.00308\%$ & $0.0573\%$\\
$1.25$ & $0.382000-0.095230 i$ & $0.382053-0.095252 i$ & $0.0137\%$ & $0.0228\%$\\
$1.45$ & $0.382485-0.095087 i$ & $0.382462-0.095071 i$ & $0.00585\%$ & $0.0169\%$\\
\hline
\end{tabular}
\caption{Comparison of the quasinormal frequencies obtained by the time-domain integration and the 6th order WKB approach with Padé approximants for $s=1/2$, $\ell=3/2$ ($M=1$).}\label{check4}
\end{table*}

\begin{table*}
\begin{tabular}{c c c c}
\hline
$\gamma $ & Frobenius & time-domain & WKB Padé\\
\hline
$0$ & $0.248263 - 0.0924877 i$  & $0.248266-0.092499 i$ & $0.248255-0.092497 i$ \\
$0.25$ & $0.248683 - 0.0924026 i$  & $0.248686-0.092415 i$ & $0.248661-0.092418 i$\\
$0.5$ &  $0.249117 - 0.0922665 i$  & $0.249120-0.092279 i$ & $0.249086-0.092290 i$\\
$0.75$ &  $0.24955 - 0.092056 i$  & $0.249554-0.092069 i$ & $0.249510-0.092084 i$ \\
$1.$ & $0.249942 - 0.0917343 i$  & $0.249948-0.091747 i$ & $0.249841-0.091792 i$ \\
$1.25$ &  $0.250138 - 0.091277 i$ & $0.250143-0.091289 i$ & $0.249969-0.091528 i$\\
$1.45$ & $0.25002 - 0.0912071 i$ & $0.250027-0.091218 i$ & $0.250064-0.091475 i$\\
\hline
\end{tabular}
\caption{Comparison of the quasinormal frequencies obtained by the Frobenius method, time-domain integration and the 6th order WKB approach with Padé approximants for $s=1$, $\ell=1$ ($M=1$).}\label{check5}
\end{table*}

\begin{table*}
\begin{tabular}{c c c c}
\hline
$\gamma $ & Frobenius  & time-domain & WKB6 Padé  \\
\hline
$0$ &  $0.457596 - 0.0950044 i$ &   $0.457599-0.095002 i$ & $0.457596-0.095005 i$ \\
$0.25$ & $0.458143 - 0.0948298 i$ & $0.458146-0.094828 i$ & $0.458144-0.094832 i$ \\
$0.5$ & $0.458692 - 0.0946179 i$ & $0.458695-0.094616 i$ & $0.458694-0.094621 i$ \\
$0.75$ & $0.459232 - 0.0943635 i$  &$0.459235-0.094361 i$ & $0.459234-0.094368 i$\\
$1.$ &  $0.459741 - 0.0940689 i$  &$0.459744-0.094067 i$ & $0.459747-0.094080 i$ \\
$1.25$ &  $0.460196 - 0.093769 i$  &$0.460199-0.093767 i$ & $0.460223-0.093782 i$ \\
$1.45$ &  $0.460571 - 0.0935554 i$  &$0.460586-0.093553 i$ & $0.460593-0.093547 i$ \\
\hline
\end{tabular}
\caption{Comparison of the quasinormal frequencies obtained by the Frobenius method, time-domain integration and the 6th order WKB approach with Padé approximants for $s=1$, $\ell=2$ ($M=1$).}\label{check6}
\end{table*}

\begin{figure*}
\resizebox{\linewidth}{!}{\includegraphics{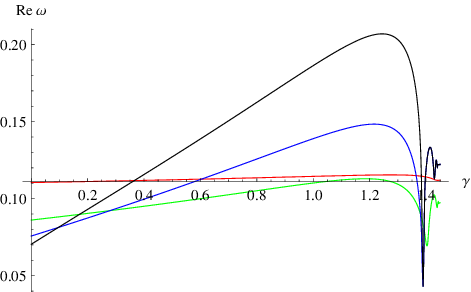}\includegraphics{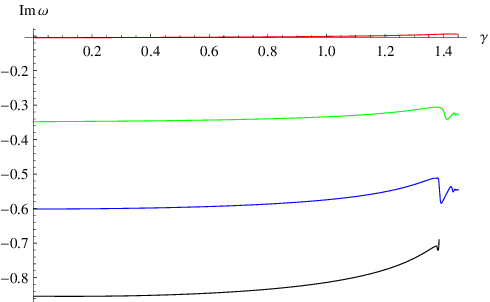}}
\caption{The fundamental mode (red, top on the left plot) and the first three overtones (green, blue and black) computed by the Frobenius method for the $\ell=0$ scalar field perturbations as a function of $\gamma$.}\label{fig:Outburst}
\end{figure*}

\subsection{WKB method}

When the effective potential $V(r)$ in the wave-like equation (\ref{wave-equation}) takes the form of a barrier with a single peak and decaying at both asymptotic regions (the event horizon and infinity), the WKB formula is suitable for obtaining the low-lying quasinormal modes that satisfy the boundary conditions.
The WKB method relies on matching the asymptotic solutions, which adhere to the quasinormal boundary conditions (\ref{boundaryconditions}), with the Taylor expansion around the peak of the potential barrier. The first-order WKB formula serves as the eikonal approximation and is exact in the limit $\ell \to \infty$. Subsequently, the general WKB expression for the frequencies can be expressed as an expansion around the eikonal limit, as follows \cite{Konoplya:2019hlu}:
\begin{eqnarray}\label{WKBformula-spherical}
\omega^2&=&V_0+A_2(\K^2)+A_4(\K^2)+A_6(\K^2)+\ldots\\\nonumber&-&\imo \K\sqrt{-2V_2}\left(1+A_3(\K^2)+A_5(\K^2)+A_7(\K^2)\ldots\right),
\end{eqnarray}
where the matching conditions for the quasinormal modes imply that
\begin{equation}
\K=n+\frac{1}{2}, \quad n=0,1,2,\ldots,
\end{equation}
with $n$ being the overtone number. Here, $V_0$ denotes the value of the effective potential at its maximum, $V_2$ represents the second derivative of the potential at this point with respect to the tortoise coordinate, and $A_i$ for $i=2, 3, 4, \ldots$ signifies the $i$th WKB order correction term beyond the eikonal approximation, dependent on $\K$ and derivatives of the potential at its maximum up to the order $2i$. The explicit form of $A_i$ can be found in \cite{Iyer:1986np} for the second and third WKB orders, in \cite{Konoplya:2003ii} for the 4th-6th orders, and in \cite{Matyjasek:2017psv} for the 7th-13th orders. This WKB approach for determining quasinormal modes and grey-body factors has been extensively utilized across various orders in numerous studies (see, for example, \cite{Abdalla:2005hu,Konoplya:2006ar,Konoplya:2006rv,Kokkotas:2010zd,Guo:2022hjp,Chen:2019dip,Fernando:2012yw,Momennia:2018hsm,Barrau:2019swg}).

\begin{figure}
\resizebox{\linewidth}{!}{\includegraphics{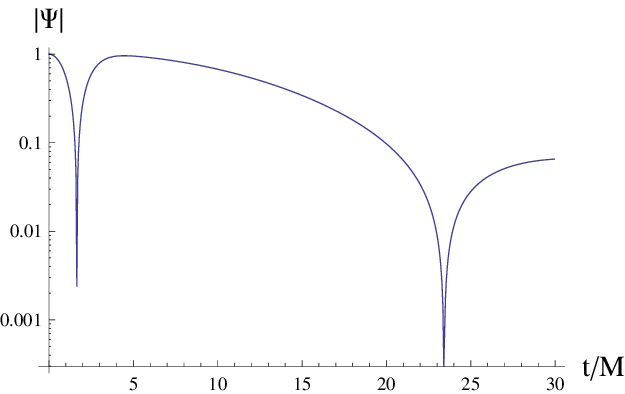}}
\caption{Example of a time-domain profile for the scalar perturbations ($\ell=0$)  hole $\gamma = 0.5$, $M =1$.}\label{fig:timedomain}
\end{figure}

\begin{figure}
\resizebox{\linewidth}{!}{\includegraphics{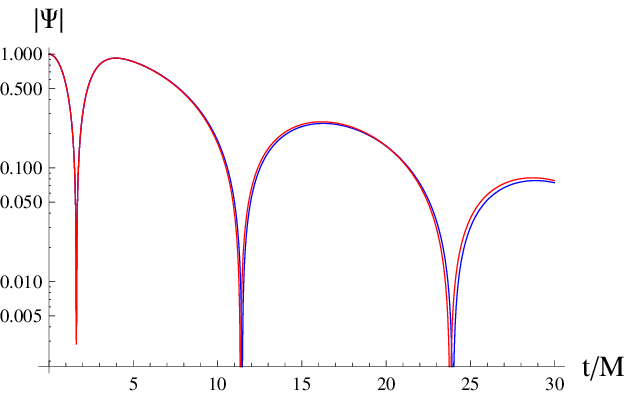}}
\caption{Example of a time-domain profile for the electromgnetic perturbations ($\ell=1$) $\gamma =0.1 $ (blue) and $\gamma =1.4$ (red); $M =1$..}\label{fig:timedomain2}
\end{figure}

\subsection{Frobenius method}

For finding precise numerical values of quasinormal frequencies we use the Frobenius method \citep{Leaver:1985ax,Leaver:1986gd}, which is based on a convergent procedure. The differential wave-like equation has a regular singular point at $r=r_0$ and the irregular one at $r=\infty$.
The new radial function $P(r, \omega)$, is introduced in such a way
\begin{equation}\label{reg}
\Psi(r)= P (r, \omega) y(r),
\end{equation}
that the factor $P(r, \omega)$ makes $y(r)$ regular everywhere in the range $r_0\leq r$ once the quasinormal modes boundary conditions are satisfied.
Then, $y(r)$ can be cast in the form of a following series:
\begin{equation}\label{Frobenius}
y(r)=\sum_{k=0}^{\infty}a_k\left(1-\frac{r_0}{r}\right)^k.
\end{equation}
Finally the Gaussian eliminations are used to reduce finding of $\omega$ to the numerical solution of a non-algebraic equation with the help of the built-in FindRoot command in {\it Mathematica}. This method was applied and discussed in a great number of publications \cite{Nollert:1993zz,Zhidenko:2006rs,Rostworowski:2006bp,Konoplya:2004uk,Zinhailo:2024jzt,Bolokhov:2023bwm,Bolokhov:2023ruj} as a precise tool for finding quasinormal modes of black holes.

\section{Quasinormal modes}\label{sec:QNM}

As the initial Calmet-Kuipers spacetime represents a wormhole, it is worth to discuss the spectra of wormholes versus black holes in this context.
The quasinormal spectra of wormholes exhibit both common and distinctive features compared to those of black holes. A primary common feature is their shared boundary conditions: in terms of the tortoise coordinate, the black hole event horizon corresponds to negative infinity, akin to the "left infinity" of a wormhole spacetime, which corresponds to the distant region of our universe or another universe \cite{Konoplya:2005et,Bronnikov:2012ch}. Quasinormal modes represent the proper frequencies corresponding to a response to a momentary perturbation, occurring when the source of the perturbation ceases. Thus, no incoming waves are required from either plus or minus infinities for either a black hole or a wormhole.

This characteristic of wormhole quasinormal modes could be leveraged to mimic the ringdown of black holes if the wormhole metric behaves similarly to a black hole, such as Schwarzschildian, across most of its space, except for a small region where the throat replaces the event horizon \cite{Damour:2007ap}. In this scenario, the only discernible difference would be a slight modification of the signal, known as echoes, at late times \cite{Cardoso:2016rao,Bueno:2017hyj,Bronnikov:2019sbx}. However, this pertains to a single dominant mode, and considering a set of frequencies, wormholes and black holes could, at least in principle, be distinguished \cite{Konoplya:2016hmd}, and the shape of the wormhole could be deduced from its spectrum \cite{Konoplya:2018ala,Volkel:2018hwb}. Quasinormal modes of various wormhole models have been extensively studied in numerous works \cite{Churilova:2021tgn,Konoplya:2010kv,Blazquez-Salcedo:2018ipc,Oliveira:2018oha,DuttaRoy:2019hij,Ou:2021efv,Azad:2022qqn,Zhang:2023kzs}, exhibiting many similar features to black holes, including the presence of arbitrarily long-lived modes \cite{Churilova:2019qph}, quasi-resonances, and an analogue of the null geodesics eikonal quasinormal modes correspondence \cite{Jusufi:2020mmy}.

Next, we will compute the quasinormal modes of the quantum-corrected black hole spacetime obtained from the wormhole metric by expanding the $B(r)$ function in terms of small $\gamma$. However, during the initial ringdown phase, the time-domain integration of the original Calmet-Kuipers wormhole metric under small values of $\gamma$ is almost indistinguishable from that of the black hole.

The quasinormal modes are calculated here using three independent methods: the Frobenius method,  the time-domain integration, and the 6th order WKB approach combined with the $\tilde{m}=4$ Padé approximants, where $\tilde{m}$ defines the structure of the Padé approximants and is defined in \cite{Konoplya:2019hlu}.
Examples of the time-domain profile for the $\ell =0$ scalar and $\ell=1$ electromagnetic  perturbations are given in figs. \ref{fig:timedomain} and  \ref{fig:timedomain2}.
From tables \ref{check1}- \ref{check6} we can see that the 6th order WKB method with the Padé approximants are in a reasonable concordance with the time-domain integration.
However, the time-domain integration is closer to the precise results given by the Frobenius method, especially for $\ell=0$ scalar perturbations.

Taking the Frobenius data as accurate for the lowest multipole numbers we can see that for all cases of $\ell>0$ the relative error of the WKB method and time-domain integration does not exceed a small fraction of one percent, being mostly smaller than $\sim 0.1 \%$ while the overall deviation of the frequency from its Schwarzschild limit reaches a few percents, which is at least one order larger than the expected relative error. Thus, we conclude that the WKB data could be trusted for $\ell>0$, while for $\ell=0$ we can rely upon the precise results of the Frobenius method and even, as reasonable approximation on time-domain integration which is based on the convergent procedure as well. The effective potential for Dirac perturbations does not have a polynomial form, so the Frobenius method cannot be directly applied. However, even in the most challenging case of $\ell = 0$ perturbations, the time-domain integration yields sufficient accuracy, with errors significantly smaller than the effect being observed. Consequently, we can rely on the time-domain integration results for $\ell = 1/2$ and higher Dirac perturbations, as the frequencies are extracted with greater precision for higher $\ell$ values, given that the ringing period is longer for larger $\ell$.
Unlike the time-domain integration, the WKB series converges only asymptotically.
From tables  \ref{check1}- \ref{check6} we see that when the quantum correction is tuned on, $Re \omega$ is almost unchanged, while the damping rate, proportional to $Im \omega$ is decreased by a few percents for the lowest multipoles.

Finally, we observe that while the fundamental mode deviates only mildly from its Schwarzschild limit — by about $10\%$ for the damping rate and less than $2\%$ for the real oscillation frequency at $\ell=0$ scalar perturbations (see Table I) — the overtones show much stronger deviations that increase with $n$. As shown in Fig. \ref{fig:Outburst}, the variation in $Re (\omega)$ for the first overtone already reaches tens of percent, while the deviation for the third overtone from its Schwarzschild limit exceeds $100\%$. This effect, sometimes referred to as "the sound of the event horizon" \cite{Konoplya:2023hqb}, is related to the high sensitivity of the first few overtones to even minimal deformations in the near-horizon zone, as described in \cite{Konoplya:2023hqb,Konoplya:2022pbc} and studied across various black hole configurations (see, for instance, \cite{Konoplya:2022hll,Bolokhov:2023bwm,Bolokhov:2023ruj,Zhang:2024nny,Zinhailo:2024kbq}). Such deformations in the near-horizon zone, which vanish in the far zone, are typical for quantum-corrected black hole solutions, as the usual post-Newtonian behavior must prevail in the far zone.

Using the expansion in terms of the inverse multipole number, one can obtain the analytic formula for quasinormal modes in the regime of large $\ell$, as it was done in a number of works for various spacetimes (for example, in \cite{Zinhailo:2019rwd,Paul:2023eep,Davey:2023fin,Konoplya:2001ji,Zhidenko:2008fp,Konoplya:2005sy,Bolokhov:2024ixe,Malik:2024sxv,Malik:2024voy,Malik:2023bxc,Malik:2024nhy}). Here, using, in addition, expansion into small $\gamma$ we obtain the position of the maximum of the effective potential $r_{m}$,
\begin{widetext}
\begin{equation}
r_{m} = \left(3 M-\frac{5 \gamma }{162 M^3}+O\left(\gamma
   ^2\right)\right)+\frac{O\left(\gamma ^2\right)}{\kappa
   }+O\left(\frac{1}{\kappa }\right)^2.
\end{equation}
Then, using the first order WKB formula, we obtain the frequency in the eikonal regime $\ell \gg n$,
\begin{equation}
\omega = \kappa  \left(\frac{1}{3 \sqrt{3} M}+\frac{5 \gamma }{4374
   \sqrt{3} M^5}+O\left(\gamma
   ^2\right)\right)+\left(-\frac{i K}{3 \sqrt{3} M}+\frac{13
   i \gamma  K}{4374 \sqrt{3} M^5}+O\left(\gamma
   ^2\right)\right)+O\left(\frac{1}{\kappa
   }\right).
\end{equation}
\end{widetext}
Here we used the following designations: $\kappa = \ell + (1/2)$ and $K =n+(1/2)$. When $\gamma =0$, the above expressions are reduced to those well-known for the Schwarzschild black hole. One can easily see that the above expressions is in concordance with the null geodesics/eikonal quasinormal modes correspondence \cite{Cardoso:2008bp}, despite a number of exceptions
described in \cite{Bolokhov:2023dxq,Konoplya:2022gjp,Konoplya:2017wot}. Indeed, from the above expressions one can deduce the rotational frequency and the Lyapunov exponent at the unstable circular null geodesics, according to the formalism described in \cite{Cardoso:2008bp}.

\section{Conclusions}\label{sec:conclusions}

We have demonstrated that the original Calmet-Kuipers spacetime does not represent a black hole but rather a wormhole. By retaining only the linear term in one of the metric functions, this spacetime can be transformed into a quantum-corrected black hole.

In this study, we computed low-lying quasinormal modes for scalar, electromagnetic, and Dirac perturbations of the quantum-corrected black hole inspired by the Calmet-Kuipers spacetime. A comparison among the precise Frobenius method, time-domain integration and the 6th order WKB method with Padé approximants reveals reasonable agreement for all cases, except for $\ell=0$ scalar perturbations, where the time-domain integration should yield reliable results. In the eikonal regime we produced the analytic formula for quasinormal modes.

Our analysis indicates that the real oscillation frequency of the fundamental mode is nearly unaffected by the quantum parameter, while the damping rate is moderately  reduced. On the contrary, the first few overtones deviate from their Schwarzschild values at a much stronger and increasing with $n$ rate. This behavior of the overtones is  similar to the one observed in \cite{Konoplya:2022hll} and closely connected to the near-horizon effect of the quantum correction.

\begin{acknowledgments}
The author wishes to thank R. A. Konoplya for useful discussions. The author acknowledges the University of Seville for their support through the Plan-US of aid to Ukraine.
\end{acknowledgments}

\bibliographystyle{unsrt}
\bibliography{bibliographyCK}

\end{document}